\documentstyle[11pt,epsfig,rotating]{article}
\newlength{\dinwidth}
\newlength{\dinmargin}
\begin{document}
\setlength{\unitlength}{1mm}
\begin{titlepage}
\begin{flushleft}
%
%===> Change report numbers and date
%
{\tt hep-ex/9504004}\\
{\tt DESY 95-062    \hfill    ISSN 0418-9833} \\
{\tt April 1995}                  \\
\end{flushleft}

\vspace*{4.cm}
\begin{center}
\begin{Large}

{\bf Inclusive Parton Cross Sections in Photoproduction \\
     and Photon Structure} \\

\vspace{1.cm}
{H1 Collaboration}    \\
\end{Large}
%================================abstract===============================
\vspace*{4.cm}
{\bf Abstract:}
\end{center}
\begin{quotation}
\renewcommand{\baselinestretch}{1.0}\large\normalsize

Photoproduction of 2-jet events is studied with the H1 detector at HERA.
Parton cross sections are extracted from the data by an unfolding method
using leading order parton-jet correlations of a QCD generator.
The gluon distribution in the photon is derived in the fractional momentum
range
$0.04\le x_\gamma \le 1$
at the average factorization scale $75$ GeV$^2$.

\renewcommand{\baselinestretch}{1.2}\large\normalsize

\end{quotation}
%\begin{center}
%\vfill
%
%===> Change the reference to the journal and the date
%
%Submitted to Journal, dd Month 1993
%      \cleardoublepage
%\end{center}
\end{titlepage}
%============================authors===========================================
\begin{Large} \begin{center} H1 Collaboration \end{center} \end{Large}
\begin{flushleft}
%MACRO 'I01NIE.H1PUB(H1AUTS)'
%\input{lo.h1auts.tex}
%  Comments to F.Niebergall       %AAA9                     AaAAAAA9
%% I.~Abt$^{7}$,                  %DAVI-LEFT   6/93         Abt
 T.~Ahmed$^{3}$,                  %BIRM-LEFT  1/2/94        Ahmed
 S.~Aid$^{13}$,                   %HAM2-PD      8/93        Aid
 V.~Andreev$^{24}$,               %LPI-PD                   Andreev
 B.~Andrieu$^{28}$,               %ECPL-PD                  Andrieu
 R.-D.~Appuhn$^{11}$,             %DESY-PD     4/92         Appuhn
% C.~Arnault$^{27}$,              %ORSA-TP                  Arnault
 M.~Arpagaus$^{36}$,              %ZUTH-ST                  Arpagaus
 A.~Babaev$^{26}$,                %MPIM-PD                  Babaev
 J.~Baehr$^{35}$,                 %ZEUT-PD                  Baehr
%% H.~B\"arwolff$^{35}$,          %ZEUT-LEFT                Baerwolff
% E.~Banas$^{6}$,                 %CRAC-TP    6/93          Banas
 J.~B\'an$^{17}$,                 %KOSI-PD                  Ban1
 Y.~Ban$^{27}$
% ...
\end{flushleft}
\begin{flushleft} {\it
%MACRO 'I01NIE.H1PUB(H1INST)'
%\input{lo.h1inst.tex}
%     H1 Institutes as appearing on publications
% ...
$\:^2$ III. Physikalisches Institut der RWTH, Aachen, Germany$^ a$ \\
 $\:^3$ School of Physics and Space Research, University of Birmingham,
                             Birmingham, UK$^ b$\\
% ...
$ ^{11}$ DESY, Hamburg, Germany$^a$ \\
% ...
$ ^{13}$ II. Institut f\"ur Experimentalphysik, Universit\"at Hamburg,
                                     Hamburg, Germany$^ a$  \\
% ...
$ ^{17}$ Institute of Experimental Physics, Slovak Academy of
                Sciences, Ko\v{s}ice, Slovak Republic$^ f$\\
% ...
$ ^{24}$ Lebedev Physical Institute, Moscow, Russia$^ f$ \\
% ...
$ ^{26}$ Max-Planck-Institut f\"ur Physik,
                                            M\"unchen, Germany$^ a$\\
 $ ^{27}$ LAL, Universit\'{e} de Paris-Sud, IN2P3-CNRS,
                            Orsay, France\\
 $ ^{28}$ LPNHE, Ecole Polytechnique, IN2P3-CNRS,
                             Palaiseau, France \\
% ...
$ ^{35}$ DESY, Institut f\"ur Hochenergiephysik,
                              Zeuthen, Germany$^ a$\\
 $ ^{36}$ Institut f\"ur Teilchenphysik,
          ETH, Z\"urich, Switzerland$^ i$\\
% ...
$ ^a$ Supported by the Bundesministerium f\"ur
                                  Forschung und Technologie, FRG
 under contract numbers 6AC17P, 6AC47P, 6DO57I, 6HH17P, 6HH27I, 6HD17I,
 6HD27I, 6KI17P, 6MP17I, and 6WT87P \\
 $ ^b$ Supported by the UK Particle Physics and Astronomy Research
 Council, and formerly by the UK Science and Engineering Research
 Council \\
% ...
$ ^f$ Supported by the Deutsche Forschungsgemeinschaft\\
% ...
$ ^i$ Supported by the Swiss National Science Foundation\\

   } \end{flushleft}

%============================text===========================================
%
\newpage
\section{Introduction}
\noindent
The interaction of electrons and protons at the HERA collider is
dominated by photoproduction processes:
electrons scatter through small angles
and emit quasi-real photons, which then interact with the protons.
The center of mass (CMS) energies in the $\gamma p$ system
reach up to 300 GeV.
A fraction of these events has large transverse energy in the final
state including the formation of jets,
as has been reported in several recent publications \cite{h1gp,H1old,zsgp}.

The jet production can be well described in the framework of QCD.
In this picture the photon couples either directly to a parton of the proton,
or indirectly via the photon's own parton content.
The first are called {\it direct} processes, which include the
QCD-Compton (Fig.1a) and photon-gluon fusion diagrams,
while the latter are usually refered to as {\it resolved} processes. An
example is shown in Fig. 1b.

\begin{figure}[htbp]
\begin{center}
\epsfig{file=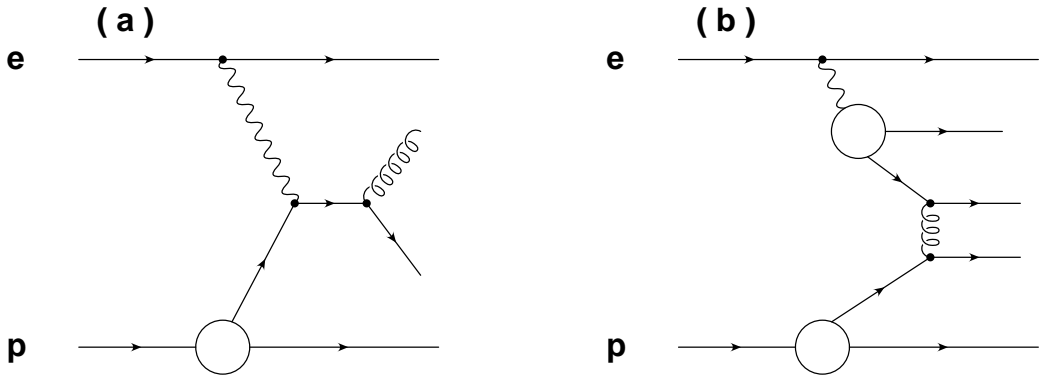,height=5cm}
\end{center}
\caption[$feynman]{\label{feynman}
Examples of diagrams for direct (a) and resolved photon (b)
processes in electron-proton scattering.}
\end{figure}
Predictions for the jet cross section are usually obtained
in leading order (LO) QCD by convoluting
the parton densities in the photon and in  the proton with hard partonic
scattering cross sections calculated at the tree level.
The partons leaving the
hard scattering reaction are identified with hadronic jets.
Due to the reduced center of mass energy of the
resolved reactions jet production from these processes could not be studied
in detail in previous fixed target photoproduction experiments.
At HERA energies however, the results of several authors~\cite{loqcd} agree in
predicting that the photoproduction of jets is
dominated by resolved reactions at low and medium
transverse energies $E_t^{jet}$ of the jets, say below $30$ GeV.
Thus, in the framework of this QCD picture,
measurement of the jet cross section can be used to
obtain information on the photon's parton content
assuming the parton densities in the proton are known.
{}From $e\gamma$ deep inelastic scattering experiments, studied
at $e^+e^-$ colliders, the quark content of the photon is already
relatively well known in the fractional momentum range of the parton
$0.007\le x_\gamma\le 1$~\cite{opal}. The photoproduction of jets
at high energies
therefore offers a new tool for the determination of the gluon density
in the photon.
Recent experiments which study jet production in $\gamma\gamma$ scattering
are also sensitive to this quantity~\cite{gamgam}.

The present analysis studies 2-jet production with more than $7$ GeV
transverse energy $E_t^{jet}$ per jet in photon-proton scattering.
The scattered electron is tagged at    small angles so   that the
photon is almost real and the energy of the
photon is known from $E_\gamma=E_{beam}-E_{tag}$.
Inclusive differential cross sections $d\sigma/dp_t$ and
$d\sigma/d\eta$ are derived at the leading order parton level.
Here $p_t$ describes the transverse momenta of the partons
and $\eta$ their pseudo-rapidities in the laboratory system.
The unfolding method described below extracts partonic cross sections
from the measured jet distributions
utilizing a Monte Carlo model which besides the LO QCD model describes
the influence of initial and final state parton showers,
multiple parton interactions, hadronization and detector effects.

The results can be directly compared to LO calculations
using different parametri\-zations for the parton densities.
In next to leading order (NLO) QCD a jet algorithm has
to be introduced at the parton level as well. The calculations presented
so far~\cite{hoqcd,kramer} take a cone algorithm and find differences
between LO and NLO predictions of the order of only $10\%-30\%$ for
values of transverse jet energy, jet pseudo-rapidity
and other jet parameters used in this paper.
%Thus, with some caution, it is possible to compare the experimental results
%with NLO predictions as well.

Besides comparing the inclusive jet cross sections with theoretical
predictions, the gluon density in the photon can also
be derived in a more direct way.
The momentum fraction $x_\gamma=E_{(parton/\gamma)}/E_\gamma$
can be fully reconstructed knowing the energies and angles of the jets and
the energy of the incoming photon.
By unfolding the measured $x_\gamma^{vis}$ distribution to
the parton level using the same Monte Carlo model mentioned above
and correcting for the quark contribution and the direct photon processes,
it is therefore possible to
determine the gluon density distribution in leading order.

The paper is structured as follows:
after a short description of the detector and of the event selection,
the Monte Carlo model used for comparisons with the data is studied
with respect to the energy flow around jets.
The model is then used to extract from the jets observed in the data
inclusive parton cross sections at the leading order QCD level which can
be directly compared to analytical QCD calculations.
It is further used to determine a distribution of the fractional
momentum $x_\gamma$ from 2-jet events which is interpreted
in terms of
a)~the direct photon contributions,
b)~the resolved contributions with a quark from the photon, and
c)~the resolved contributions with a gluon from the photon.

\section{Detector Description and Selection of 2-jet Events}

A detailed description of the H1 apparatus can be found elsewhere~\cite{H1NIM}.
The following briefly describes the components of the detector relevant
for  this analysis, which makes use of the calorimeters,
the luminosity system and the central tracking detector.

The LAr calorimeter~\cite{LARC}  extends over the polar angular range
$4^\circ < \theta <  153^\circ$ with full azimuthal coverage, where
 $\theta$ is defined with respect to the proton
beam direction ($+z$ axis).
The  calorimeter  consists   of an electromagnetic section with
lead absorbers, corresponding to a depth of between 20 and 30 radiation
lengths, and a hadronic section with steel absorbers.
The total depth of the LAr~calorimeter varies between 4.5 and 8
hadronic interaction lengths. The calorimeter is highly segmented in
both sections  with a total of around 45000 cells.
The electronic noise per channel is typically in the range 10 to 30
MeV (1 $\sigma$ equivalent energy).
Test beam measurements of LAr~calorimeter modules have demonstrated
energy resolutions of $\sigma(E)/E\approx 0.12/\sqrt{E}\oplus 0.01$
with $E$ in GeV for electrons~\cite{H1joerg}
and $\sigma(E)/E\approx 0.5/\sqrt{E}\oplus 0.02$
for charged pions~\cite{H1NIM,H1PI}.
The hadronic energy scale and resolution have been
verified from the balance of transverse momentum between hadronic
jets and the scattered electron in deep inelastic scattering events
and are known to a precision of $5\%$ and $10\%$ respectively.

The calorimeter is surrounded by a superconducting solenoid providing a uniform
magnetic field of $1.15$ T parallel to the beam axis in the tracking region.
Charged particle tracks are measured in two concentric jet drift chamber
modules (CT), covering the polar angular range $ 15^\circ < \theta <
165^\circ$,
and a forward tracking detector (FT), covering the
range $ 7^\circ < \theta < 25^\circ$.

The luminosity system consists of two TlCl/TlBr crystal calorimeters having
a resolution of $\sigma(E)/E=0.1/\sqrt{E}$ with $E$ in GeV. The
electron tagger is located at $z=-33$ m and the photon
detector at $z=-103$ m. The electron tagger accepts electrons with an energy
fraction between 0.2 and 0.8 with respect to the beam energy
and  scattering angles below $\theta'\le 5$ mrad $(\theta' = \pi -\theta)$.

The events used in this analysis were taken during the 1993 running period,
in which HERA collided
26.7 GeV electrons on 820 GeV protons, and correspond to an integrated
luminosity of $290$ nb$^{-1}$.
They were triggered by a coincidence of the electron tagger
and at least one track from the central jet chamber trigger.
Events were selected, if they fullfilled the following criteria:
\begin{enumerate}
\item The energy deposited in the electron tagger was in the range
      $8 \le E_{tag} \le 20$ GeV.
      The cross sections refer to a
      scaled photon energy of $0.25\le y \le 0.7$ and
      a negative squared four-momentum of the photon of $Q^2\le 0.01$ GeV$^2$.
      For the sample used to determine the inclusive parton cross sections
      an additional containment cut for the electron shower was applied
      in order to facilitate the acceptance calculation of the electron
      tagger.
%tagger of $\vert x_{tag} \vert \le 6.5$ cm.
\item At least one track in the central tracker with transverse momentum above
      $0.3$ GeV
      coming from the interaction region was required to determine the
      position of the vertex along the beam axis.
\item The width of the vertex distribution along the beam axis was
      $\sigma =10$ cm.
      Events were accepted in the region of $\pm 3$ standard deviations
      around the nominal vertex position.
\end{enumerate}
Jet reconstruction was based on purely calorimetric measurements using
a cone algorithm \cite{snow} in a grid of the azimuthal angle
$\varphi^{cell}$ and pseudo-rapidity $\eta^{cell}$ which extends from
$-3\le\eta^{cell}\le 3$.
%At large rapidities $2\le\eta^{jet}\le 2.5$ the observed jets get
%%approximately
%1 unit wide in rapidity and azimuth due to scattering effects in the
%detector.
%Therefore
The cone radius $R=\sqrt{\Delta\eta^2+\Delta\varphi^2}$ in the standard
analysis was chosen to be $R=1.0$ and
$R=0.7$ was also used for cross checks.
Jets were ordered according to the transverse energy in the cone.
Events were accepted, if
\begin{enumerate}
\item At least 2 jets were found, each with transverse energy above
      $E_t^{jet}\ge 7$ GeV.
\item The jets were contained in the LAr calorimeter
      $0 \le \eta^{jet} \le 2.5$.
\item The rapidity difference between the two most energetic jets was less than
      $\vert\Delta\eta\vert \le 1.2$
      in order to reject events where the photon spectator
      (Fig.\ref{feynman}b) is
      misidentified as a jet from the hard parton-parton scattering process.
\end{enumerate}
%The trigger efficiency for these events is $94\pm1\%$
The trigger efficiency was determined to be $94\pm1\%$ using a monitoring
trigger with full efficiency for this event selection.
A correction of the jet energy scale with respect to the Monte Carlo model
leads to an additional selection criterion which is described below.
The total number of 2-jet events remaining is 366 without the cut on the
shower containment in the electron tagger,
and 292 for all cuts.

\section{Monte Carlo Generator for QCD Processes}

For the analysis of the data, the PYTHIA 5.6 event
generator for photon-proton interactions~\cite{pythia} was used
together with a generator for quasi-real photons.
%~\cite{ijray}.
PYTHIA is based on leading order (LO) QCD matrix elements
and includes initial and final state parton shower models.
The strong coupling constant $\alpha_s$ was calculated in first order QCD using
$\Lambda_{QCD}=200$ MeV with $4$ flavours.
The renormalization and factorization scales were both set to the
transverse momentum $p_t$ produced in the parton-parton scattering.
Since the QCD calculation used here is divergent for processes with small
transverse momenta of the partons emerging from the hard interaction a
lower cut-off has been applied in PYTHIA. This was set to $p_t\ge 2$ GeV.

For the proton structure the GRV-LO~\cite{pgrv} leading order parton
density parametrizations were used.
The GRV-LO~\cite{ggrv} leading order parametrizations were used
for the photon structure.
The latter give a consistent description of the data as will be shown below.
Optionally, PYTHIA allows for additional interactions within the same event.
These are LO QCD processes between partons from the photon remnant
and partons from the proton remnant.
This so called {\it multiple interaction} option has been explored in
proton-antiproton collisions before \cite{pythia, afs} and
the same parameters have been used here.
For the hadronization process the LUND fragmentation scheme was applied
(JETSET~\cite{jetset}).

The detector response for the generated events was simulated with a
detailed simulation program and then reconstructed with the same
program as used for the data.
The generated events therefore allow the calculation of
                                      correlations between
the jets reconstructed in the detector and the underlying parton kinematics
(see below).
These correlations will be used to study the parton-parton scattering
processes with the jets observed in the data.

\section{Energy Flow in Jet Events}

The precision of the measurement of the transverse jet energy $E_t^{jet}$
and how well the jet energy correlates with the parent parton momentum
$p_t$ are critical matters.
This is because the transverse jet energy distributions fall like
$(E_t^{jet})^{-n}$ where $n\sim 5.5$~\cite{H1old} so that an imperfect
description of the energy flow around the jet direction by the
Monte Carlo model is a potential source of serious error in
the conclusions on the parton scattering process.
The fact prevented relevant conclusions on the photon structure
in previous publications \cite{h1gp,H1old,zsgp}
and is taken into account in this paper for the first time at HERA.

%The measurement of the transverse jet energy $E_t^{jet}$ and
%its correlation to the parton momentum $p_t$ is very critical
%since the transverse jet energy distributions fall like $(E_t^{jet})^{-n}$
%where $n\sim 5.5$ \cite{H1old}.
%An imperfect description of the energy flow around the jet
%direction by the Monte Carlo model will most likely cause errors in the
%conclusions on the parton scattering process.

\begin{figure}[tbp]
\begin{center}
\epsfig{file=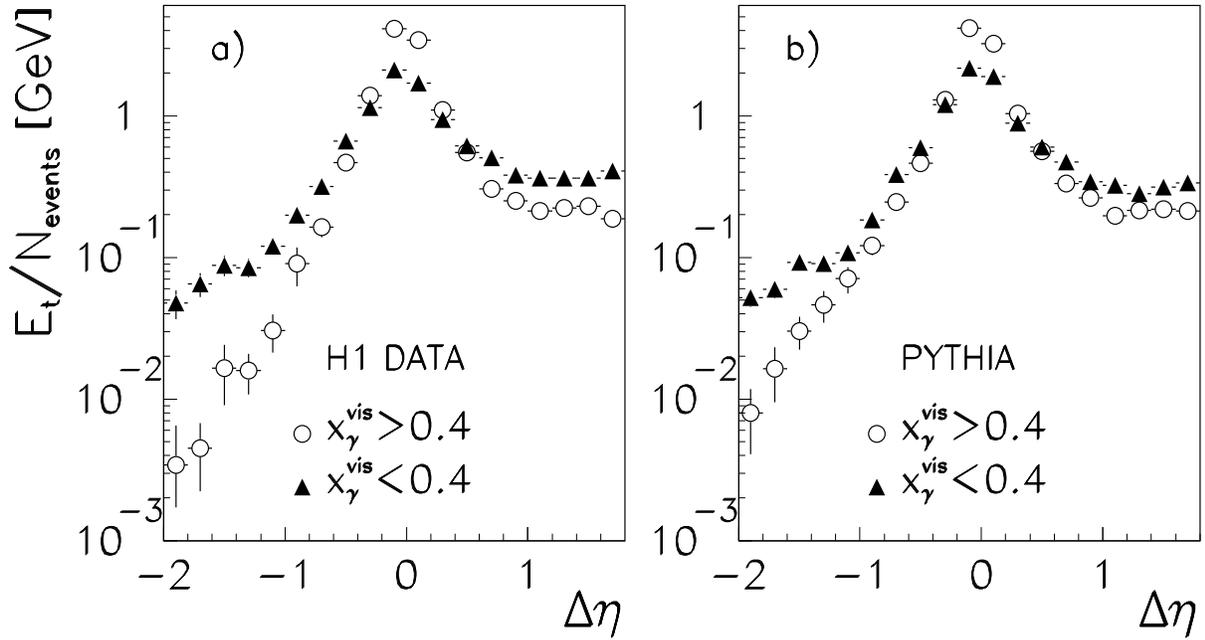,width=16cm,height=9cm}
\end{center}
\caption [$etaflow] {\label{etaflow}
Observed transverse energy flow versus the rapidity distance from
the jet direction (integrated over
$\vert\varphi^{cell}-\varphi^{jet}\vert\le 1$)
for a) data and b) PYTHIA with multiple interactions.
The jets were selected with transverse energy $7\le E_t^{jet}\le 8$ GeV
in a cone of size $R=1$ and jet rapidities between $0\le\eta^{jet}\le 1$.
The energy flow depends on the momentum fraction $x_\gamma$ of the parton
from the photon: shown are large $x_{\gamma}^{vis}\ge 0.4$ (open circles)
and small $x_{\gamma}^{vis}\le 0.4$ (filled triangles).}
\end  {figure}

In Fig.\ref{etaflow}a the observed transverse energy flow around the jet
direction per event is shown versus the rapidity distance from the jet axis
in a slice of $\vert\varphi^{cell}-\varphi^{jet}\vert\le 1$.
As an example jets were selected with transverse energy
$7\le E_t^{jet}\le 8$ GeV collected in a cone of size $R=1$
and pseudo-rapidity between $0\le\eta^{jet}\le 1$.
The jet profiles are asymmetric, showing a higher energy level in the direction
of the proton ($\Delta\eta\ge 0$) compared to the photon direction
($\Delta\eta\le 0$).
%This can be explained by QCD radiation off the initial state partons
%which is - due to the assymmetric beam energies - larger for partons
%from the proton side.
It is interesting to note that the energy flow depends not only on
rapidity, but also on the momentum fraction $x_\gamma$ of the parton
from the photon side.
The parton momentum fraction can be reconstructed using the
two jets with the highest transverse energy $E_t^{jet}$ in the event and
their pseudo-rapidities $\eta^{jet}$ together with the energy of the
photon $E_\gamma$:
\begin{equation}
x_\gamma^{vis} = \frac{E_t^{jet1}e^{-\eta^{jet1}}+E_t^{jet2}e^{-\eta^{jet2}}}
                      {2E_\gamma}
\label{xg}
\end  {equation}
The jet profiles are shown in two bins of $x_\gamma^{vis}$,
above and below $0.4$.
%$x_\gamma^{vis}\ge 0.4$ and $x_\gamma^{vis}\le 0.4$.
Two effects are observed:
\begin{enumerate}
\item In the photon direction ($\Delta\eta\le 0$) the
      low $x_\gamma^{vis}$ data show an
      enhanced energy flow relative to the high $x_\gamma^{vis}$ data.
      This can be ascribed to the remnant particles of the photon which should
      be reduced at high $x_\gamma$, and absent altogether in the case of
      direct photon processes for which $x_\gamma=1$.
\item In the proton direction ($\Delta\eta\ge 0$) the energy
      flow is also enhanced in the low $x_\gamma^{vis}$ distribution
      compared with the high $x_\gamma^{vis}$ data.
      This increased energy flow indicates additional event activity
      for events with small parton momenta $x_\gamma$, or
      large energies $\sim(1-x_\gamma)$ of the photon spectators.
\end{enumerate}

      The jet core is narrower in the case of the high $x_\gamma^{vis}$ sample
      compared to the low $x_\gamma^{vis}$ sample.
      This effect is connected to the energy flow around the jet and
      the jet energy interval: since the jets were required to have
      $\sim 7$ GeV in the cone,
      a reduced energy flow around
      the jet enforces more energy in the core.

In Fig.\ref{etaflow}b the transverse energy flow is shown for events of
the PYTHIA model including multiple interactions.
The model also shows an excess of low $x_\gamma^{vis}$ over high
$x_\gamma^{vis}$ data at positive as well as at negative rapidities.

\begin{figure}[tbp]
\begin{center}
\epsfig{file=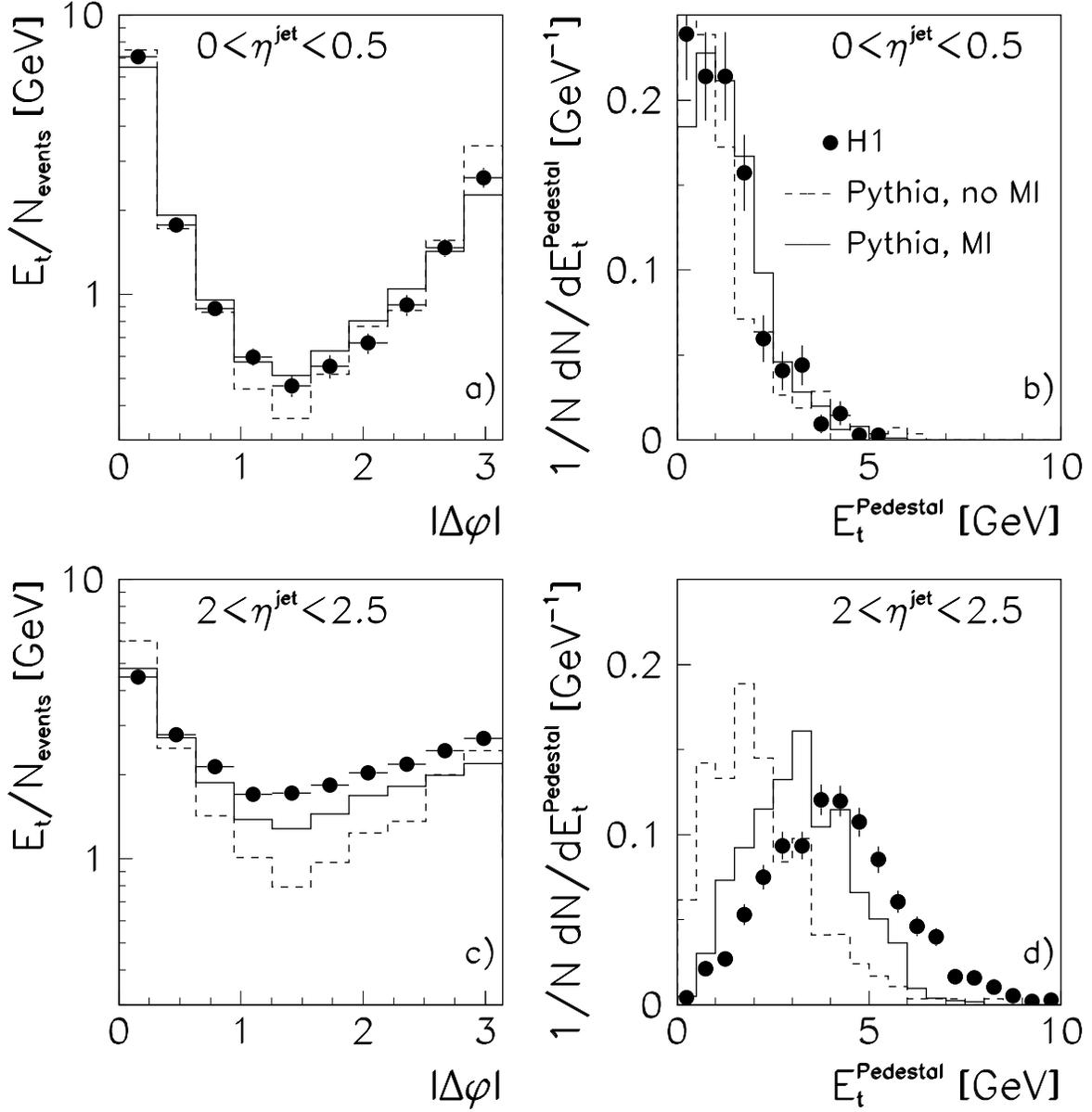,width=16cm}
\end{center}
\caption [$phiflow] {\label{phiflow}
a,c) Observed transverse energy flow versus the azimuthal angle
with respect to the jet direction
(integrated over $\vert\eta^{cell}-\eta^{jet}\vert\le 1$)
in two rapidity bins:
a) $0\le\eta^{jet}\le 0.5$ and c) $2\le\eta^{jet}\le 2.5$.
The jets have transverse energies $7\le E_t^{jet}\le 8$ GeV
in a cone of size $R=1$.
Full circles are H1 data.
The histograms refer to PYTHIA events with (full line) and
without (dashed line) multiple interactions.
b,d) Distribution of the transverse energy measured outside of the jets
in the slice $\vert\eta^{cell}-\eta^{jet}\vert\le 1$,
normalized to the area $\pi R^2$.}
\end  {figure}

The observed transverse energy flow in the azimuthal angle
$\vert\Delta\varphi\vert$
around the jet direction is shown in Fig.\ref{phiflow}a,c for two bins of the
jet rapidity.
The energy flow has been integrated in a slice of
$\vert\eta^{cell}-\eta^{jet}\vert\le 1$ around the jet axis.
The transverse energy of the jets was again restricted  to
$7\le E_t^{jet}\le 8$ GeV summed in a cone of size $R=1$.
The H1 data are shown as full circles.
The observed energy flow outside of the jet cone
is much higher for jets at large rapidities
$2\le\eta^{jet}\le 2.5$ (Fig.\ref{phiflow}c)
compared to jets in the central detector region
$0\le\eta^{jet}\le 0.5$ (Fig.\ref{phiflow}a)
and is increasing in between $0.5\le\eta^{jet}\le 2$ (not shown).
Due to the cut on the rapidity difference of the two jets
$\vert\Delta\eta\vert\le 1.2$ a part of the second jet is always seen at
$\Delta\varphi\sim\pi$.

The average transverse energy flow per cone area $\pi R^2$ determined outside
of the jet cone provides a measure of the {\it jet pedestal},
often called underlying event \cite{ellis}.
This jet pedestal is later used to estimate the amount of energy inside
the jet cone which is,
according to the PYTHIA model with multiple interactions,
not due to the fragmentation of the hard scattered partons.
The pedestal energy was determined event by event in the slice
$\vert\eta^{cell}-\eta^{jet}\vert\le 1$ and
$\vert\varphi^{cell}-\varphi^{jet}\vert\ge 1$.
Regions affected by the second jet were excluded from the measurement.
The distributions of the jet pedestals corresponding to the jet profiles of
Fig.\ref{phiflow}a,c are shown in Fig.\ref{phiflow}b,d.

The PYTHIA model without additional interactions (dashed line)
does not give a good description of the jet profiles and pedestal
distributions.
In this model the pedestal energy
corresponds to QCD radiation and fragmentation effects.
The QCD radiation effects are approximately taken into account
by the parton shower models included in PYTHIA.

The PYTHIA model with multiple interactions provides an improved
description of the jet shapes, which is reasonable in the rapidity range
$0\le\eta^{jet}\le 0.5$,
but shows deviations from the data at large jet rapidities
$2\le\eta^{jet}\le 2.5$.
%A change of some parameters connected to the multiple interaction
%option in PYTHIA did not result in an improved description of the data.
%The present analysis does not aim for a perfect tuning of a Monte Carlo model,
%but rather requires a model which can be used for corrections.
The differences between data and Monte Carlo model
are still too large to be neglected.
Moreover the sometimes large
contributions of energy due to multiple interactions in the jet cone give
a poor correlation between parton and jet energies because
this contribution leads to events entering the $E_t^{jet}\ge 7$ GeV sample
which have parton transverse momentum below the cutoff used in the
Monte Carlo model, namely $2$ GeV.

In order to avoid this problem the transverse
energies of the jets in the data were reduced by the average pedestal
difference between data and PYTHIA without multiple interactions.
In this way the jets were only corrected for the additional energy flow
over and above the initial and final state radiation and fragmentation
effects.
A corresponding
pedestal subtraction was applied to the jets of the PYTHIA events with
multiple interactions. The corrections  were parametrized in terms of
the rapidity of the jets so as to match the effect seen in the data.
The corrections vary between $0.3-2.3$ GeV for the data and $0.3-1.3$ for
the events of the PYTHIA model with multiple interactions.
This subtraction evidently also corrects for the difference in the
pedestal energies between Monte Carlo and data, provided that this
energy difference is the same on average inside and outside the cone; which
would be the case for multiple interactions. As an additional
safeguard the transverse energy of the jets is required to be above
$E_t^{jet}\ge 7$ GeV even after this jet pedestal correction which ensures
that the parent parton transverse momentum exceeds $2$ GeV.
In all figures of the following sections these jet energy corrections will
be applied to data events and Monte Carlo events.

Overall, the multiple interaction option gives an improved description
of the energy flow and offers a natural explanation of the observed
effects.
However, the improved description of the jet shape by this model
cannot be considered as a definite proof of the existence of multiple
interactions.
Therefore a model dependence remains
in this analysis where
%when
the PYTHIA version with multiple
interactions is used to extract direct information from the observed
jet spectra on the underlying (LO) parton-parton scattering processes.

\section{Jet-Parton Correlations}

\begin{figure}[tbp]
\begin{center}
\epsfig{file=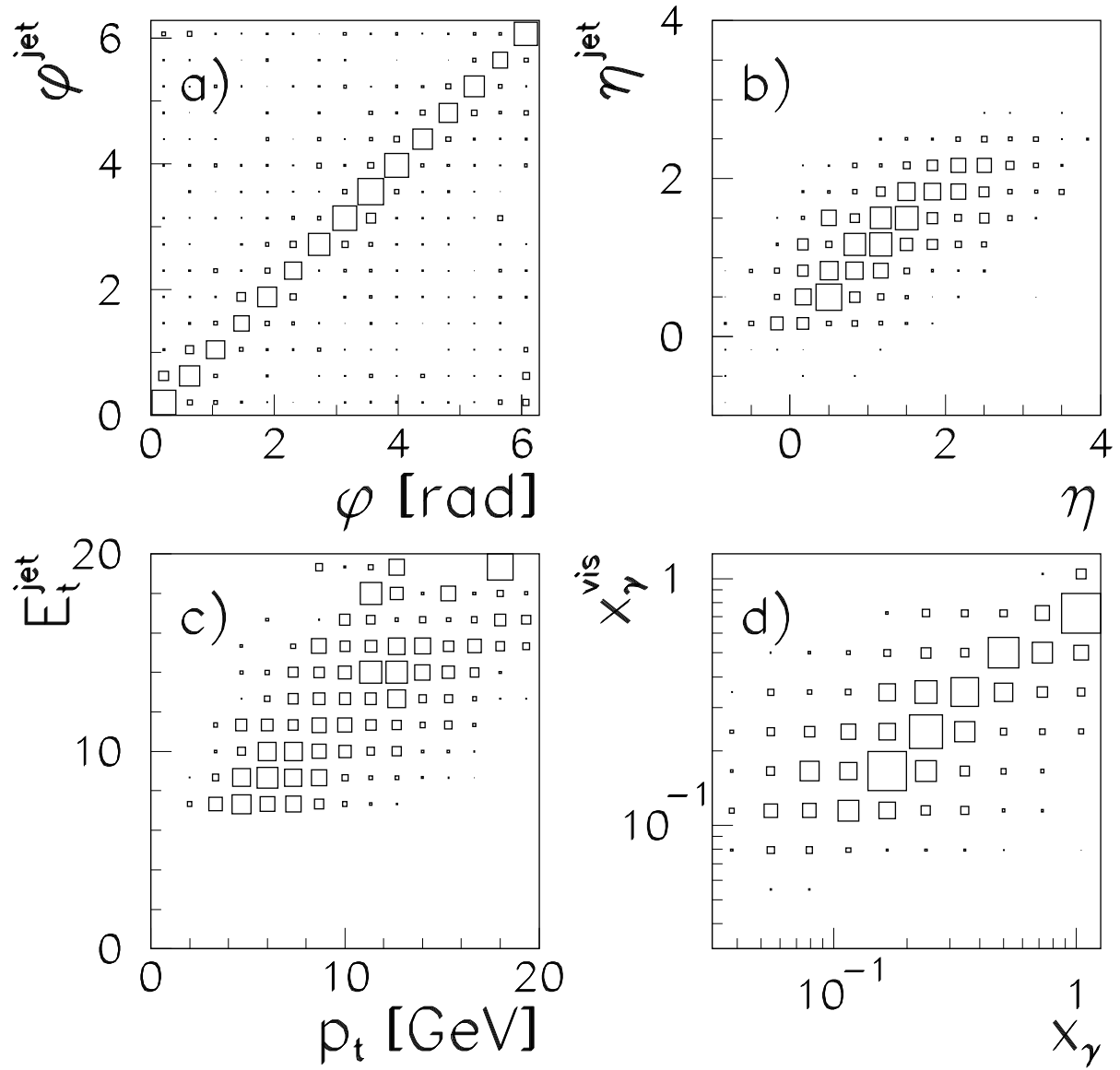,width=16cm,height=16cm}
\end{center}
\vspace{-1.2cm}
\caption [$correlation] {\label{correlation}
Correlations between reconstructed jets and leading order
parton quantities according to PYTHIA with multiple interactions:
a) azimuthal angle $\varphi$,
b) rapidity $\eta$,
c) transverse parton momentum $p_t$
  (to account for the steeply falling distribution the histogram
   has been weighted with the transverse jet energy $(E_t^{jet})^4$ ),
d) momentum fraction $x_\gamma$ of the parton from the photon side.}
\end  {figure}

The predicted correlations obtained with PYTHIA between jet and leading order
parton quantities are shown in Fig.\ref{correlation}:
a) for azimuthal angles $\varphi$;
b) for pseudo-rapidities $\eta$;
c) for the transverse jet energy $E_t^{jet}$
   and the parton transverse momentum $p_t$; and
d) for the reconstructed parton momentum fraction
from the photon side (Equation \ref{xg}) with respect to the
true $x_\gamma$ at the parton-parton scattering process.
In the leading order picture, two hard partons exist per event, and
there are at least two jets.
Therefore for the $\varphi$, $\eta$, and $p_t$ distributions
the jet with the highest $E_t^{jet}$ was
correlated with the parton giving the smallest invariant jet-parton mass,
and the second jet with the other parton.
%In all variables the jets correlate with the parton quantities.

Also, in the case of $x_\gamma^{vis}$ (Fig.\ref{correlation}d),
the two highest $E_t^{jet}$ jets were used.
%The pseudo-rapidities of the jets $\eta_i$ in the numerator
%of (\ref{xg}) were calculated from the polar angle of the jet axis.
According to the nominator of (\ref{xg}), different values of $x_\gamma^{vis}$
correspond to different 2-jet configurations:
small values of $x_\gamma^{vis}$ require two jets with
large rapidities and small transverse jet energies.
Here the remaining jet pedestal, discussed in Section 4
(dashed line of Fig.\ref{phiflow}c), raises $x_\gamma^{vis}$ relative
to the true $x_\gamma$.
Events with large $x_\gamma^{vis}$ have at least one of the jets
at small rapidity or large transverse jet energies.
At small rapidities the remaining jet pedestal is small relative to the
total $E_t^{jet}$ (dashed line of Fig.\ref{phiflow}a),
and $x_\gamma^{vis}$ corresponds in average to the true $x_\gamma$.
At $x_\gamma\sim 1$ transverse energy deposited outside of the jets reduce
$x_\gamma^{vis}$ relative to the true $x_\gamma$.
The photon energy in the denominator of (\ref{xg}) was precisely
determined from the energy of the scattered electron measured
in the electron tagger system.
The resolution in the logarithm of the reconstructed parton momentum
fraction is approximately
Gaussian in $log_{10}(x_\gamma)-log_{10}(x_\gamma^{vis})$ and varies between
$0.22$ at true parton momenta around $x_\gamma\sim 0.05$ and
$0.16$ at $x_\gamma\sim 0.5$.

{}From the study of the jet-parton correlations, the momenta of the
hard partons can be reconstructed from the measured jet energies.
It is then possible to extract partonic cross sections and
leading order parton densities.
The cross sections can be compared directly to leading order QCD calculations
instead of smearing the theoretical cross section with
transverse momenta of the partons inside the proton and the photon,
QCD radiation effects, fragmentation, detector effects, jet formation,
and then confronting it with the data.

In principle such an unfolding procedure seems to be straightforward.
The distribution of an observable $g^{det}$ measured by the detector is
related to the distribution of a
partonic observable $f^{part}$ by an integral  equation  which
expresses the convolution of the true distribution with  all  effects
%$A(u,w)$
between the creation of the hard parton and the measurement  process
$ g^{det}(u)=\int A(u,w) f^{part}(w) dw$.
This integral equation can be transformed  to  a  matrix  equation.
Solving this matrix equation
thus leads directly to the histogram $f^{part}(w)$ and
therefore e.g. to the partonic cross section.
This simple method can produce spurious oscillating components in the
result due to the limited detector resolution.
Therefore the method has to be improved by a regularization
procedure which reduces the resulting correlations by optimizing both the
number and position of the bins for the unfolded variable.
%In practice however the conversion of the integral equation to a matrix
%equation and the statistical fluctuations of the matrix elements make
%the inversion impossible and a
%regularization procedure has to be applied.

In the analysis presented here
an unfolding method \cite{Blobel} has been used to determine the differential
cross section $d\sigma/dp_t$ and to derive the distribution of the
parton momentum $x_\gamma$.
Because the resolution in the transverse momentum $p_t$
is much worse than        the resolution in rapidity $\eta$,
for the extraction of $d\sigma/d\eta$ the transverse momentum
$p_t$ is unfolded for three different bins in $\eta$.
The $p_t$ integrated results then give the
inclusive cross section $d\sigma/d\eta$.

Technically, the unfolding procedure delivers a weight  for  each Monte Carlo
event in terms of the true value of the observable
chosen for the unfolding which then can be  used
to reweight Monte Carlo distributions  of  different  variables.  The
comparison of these distributions with the data give an important check
of the transformation, as  described  in  the  Monte Carlo simulation,
between  the  partonic  distributions  and  the
measurable distributions.
An accurate Monte Carlo description of this transformation
is essential for a reliable unfolding.

\section{Parton Cross Sections}

\begin{figure}[tbp]
\begin{center}
  \epsfig{file=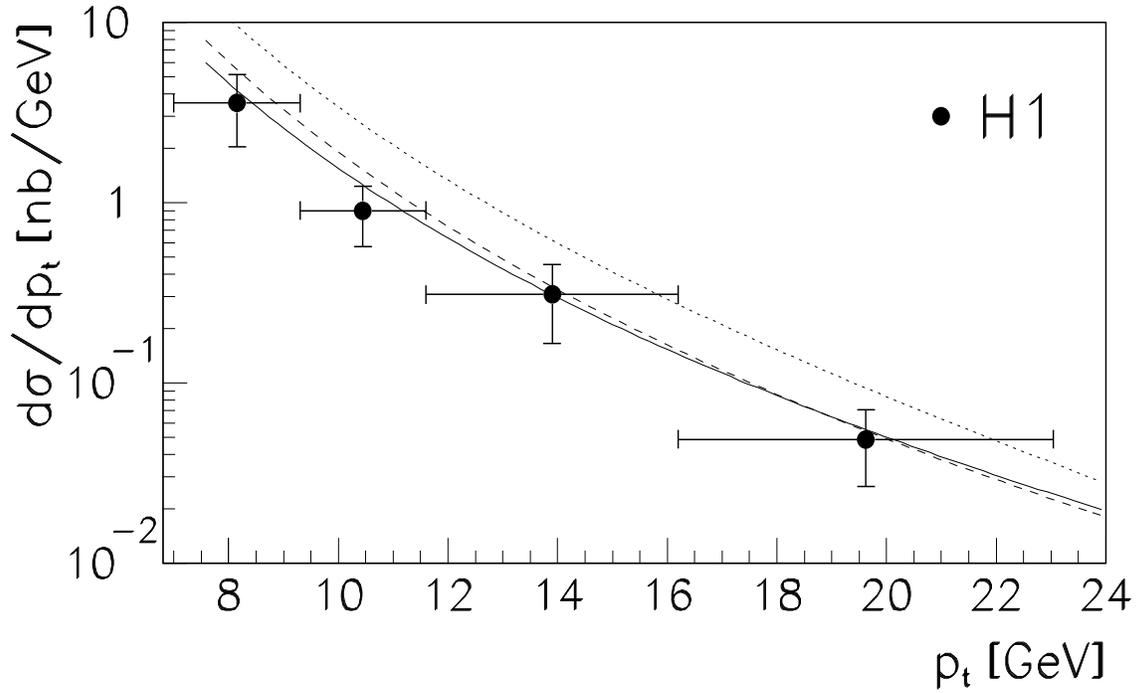,width=15cm,height=9cm}
\caption [$pT]
  {\label{pT}
Single inclusive parton $ep$ cross section unfolded to the leading order
partonic matrix element versus transverse momentum $p_t$ for
$0<\eta<2.5$. The solid line is the partonic cross section obtained
from a leading order QCD calculation \cite{salesch}
using the GRV leading order parton densities
for the proton and the photon. For the dashed (dotted) line the photon
parton densities are taken from the LAC1 (LAC3) parametrization.}
\end{center}
\end{figure}

In Fig.\ref{pT} the unfolded single inclusive parton $ep$ cross section
$d\sigma/dp_t$ is shown.
The integration over the parton rapidity $0<\eta <2.5$,
the initial photon energy and momentum transfer squared
has already been carried out.
The unfolding was done in the $p_t$ variable alone.
Migration effects in $p_t$ influence the result in the lowest $p_t$ bin
most strongly.
In order to minimize these effects
an option in the algorithm was used which allows to constrain the
cross section in the bin $4\le p_t\le 7$ GeV (not shown in the figure)
to a reasonable extrapolation of the results shown in the figure.
In addition, it ensures small bin to bin correlations (less than $30\%$) and
a smooth behaviour of the reweighting function.
%independently of the rapidity interval envisioned for the partons.
%migration effects in rapidity need to be considered.
%They are of the order of $10\%$ and have been accounted for by
%a correction factor method in the figure.
Because the $p_t$ distribution is steeply falling, the largest systematic error
stems from an error in the determination of the calorimetric energy.
Variations of the constraint on the cross section in the unseen bin
and other unfolding parameters result in a $20\%$ error on the
cross section shown in the figure.
Because the result is given in the parton rapidity range $0\le\eta\le 2.5$,
but the correlation used for unfolding is for all parton rapidities,
an additional systematic error of $25\%$ is included
to account for a possible difference in the correlations of the full
and the restricted sample.
%The systematic error in Fig.\ref{pT} has been obtained by repeating
%the unfolding with the calorimetric energy rescaled by $\pm 5\%$.
%It can be seen that the error in the three lower $p_t$ bins is completely
%dominated by systematic effects.
Other systematic error sources are described in Section 7.
The error bars are computed from the quadratic addition of statistical and
systematic errors.
The cross section, statistical and systematic errors can be found in
Table~\ref{parton_values}.

The solid line in Fig.\ref{pT} is the result of a leading order
calculation \cite{salesch} using the GRV parton densities in leading
order QCD for the proton and the photon.
The factorization and renormalization scales are
given by $p_t^2$.
The QCD parameter $\Lambda$ was set to $200$ MeV.
Leading order calculations for other photon parametrizations
(LAC1: dashed line, LAC3: dotted line \cite{lac}) are also included.
A NLO QCD calculation has been carried out by the same authors \cite{salesch}
using the GRV
higher order parton densities in the photon and the proton,
the 2-loop calculation of $\alpha_s$, $\Lambda_{\overline{MS}}^{(4)}=200$ MeV,
and a cone size of $R=1$.
Compared to the LO calculation the NLO calculation
is essentially larger by an overall factor $1.25$.
%The statistical errors on the calculations are negligible.

\begin{figure}[tbp]
\begin{center}
\epsfig{file=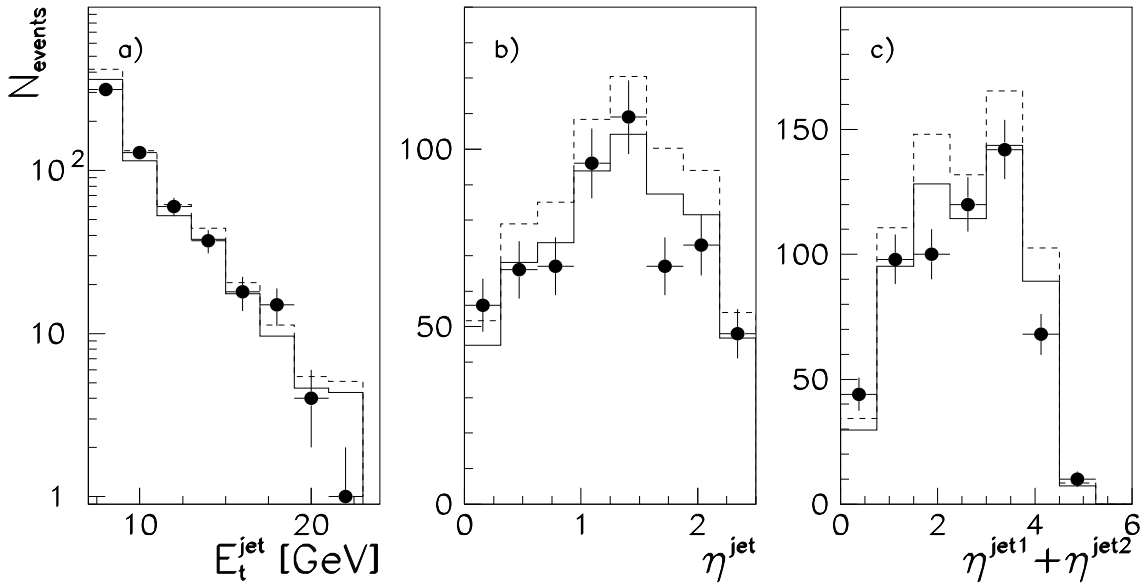,width=15cm,height=8cm}
  \caption [$check]
  {\label{check}
 Distributions for the unfolding of $p_t$.
Shown is the comparison of data
on detector level with the Monte Carlo model after (solid line) and before
(dashed line) reweighting. The parton densities are taken from the GRV
parametrizations. a) The $E_t^{jet}$ distribution of the jets,
b) the pseudo-rapidity distribution
of the jets, c) the distribution of $\eta^{jet1} +\eta^{jet2}$ for
the two highest $E_t^{jet}$ jets.
  }
\end{center}
\end{figure}

As discussed in the previous Section an important check of the
success of the transformation consists in the comparison of a set of measured
observables with the result of the reweighted Monte Carlo simulation.
This is shown in Fig.\ref{check}. The distribution
of the measured transverse jet energy $E_t^{jet}$ is
well described by the Monte Carlo simulation after reweighting (solid
line). Because the initial set of parameters already gives a good description
of the data (dashed line) the new weights differ only slightly from one.
Also the agreement between the data and the model in other variables
like $\eta^{jet}$ for all jets and $\eta^{jet1}+\eta^{jet2}$ for the two
highest
$E_t^{jet}$ jets (Figs. \ref{check} b,c) is at a satisfactory level.

\begin{figure}[tbp]
\begin{center}
\epsfig{file=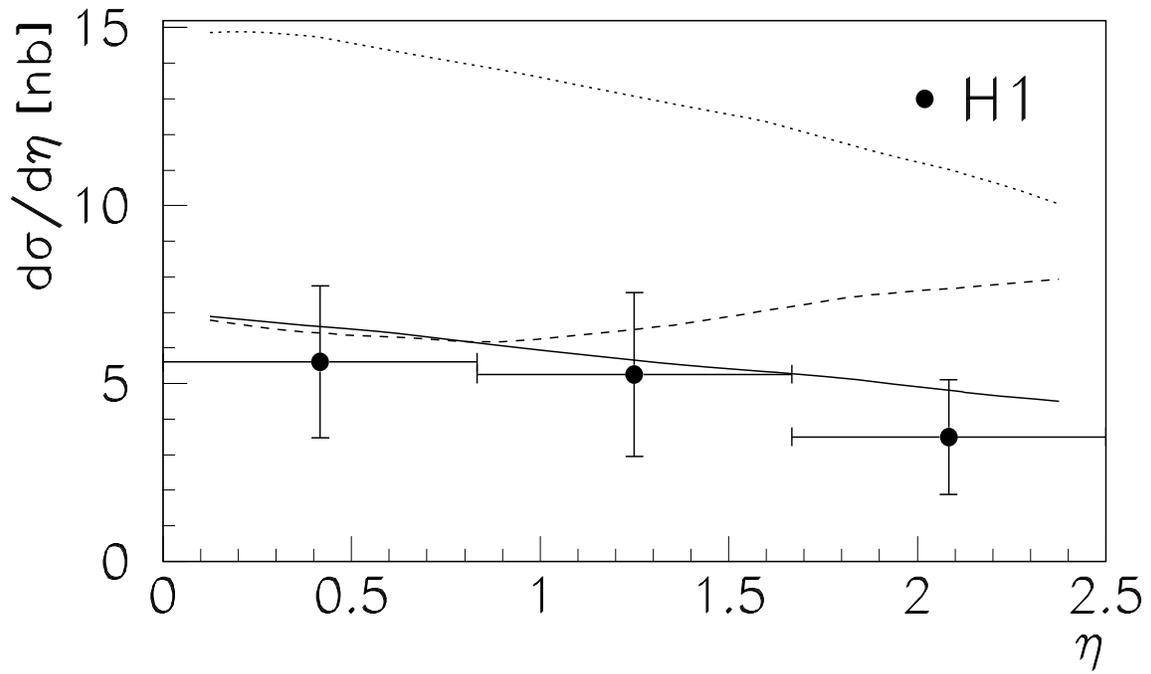,width=15cm,height=9cm}
  \caption [$eta]
  {\label{eta}
Single inclusive parton $ep$ cross section unfolded to the leading order
partonic matrix element versus pseudo-rapidity $\eta$ for
$p_t>7$ GeV. The solid line is the partonic cross section obtained
from a leading order QCD calculation \cite{salesch}
using the GRV leading order parton densities
for the proton and the photon. For the dashed (dotted) line the photon
parton densities are taken from the LAC1 (LAC3) parametrization.}
\end{center}
\end{figure}

The unfolded inclusive parton $ep$ cross section $d\sigma/d\eta$ is shown in
Fig.\ref{eta} together with the theoretical predictions. Statistical and
systematic errors are treated in the same way as in Fig.\ref{pT}.
The flat shape of the
        distribution is well reproduced by the LO QCD calculations
using the three parametrizations of the photon structure function
as above.
The absolute rate is consistent with the calculation
using the GRV parametrization.
The NLO QCD calculation again is larger by an overall factor $1.25$
with respect to the LO calculation.

\section{Parton Momentum Distributions in the Photon}

In this Section the full 2-jet kinematics is used to check the
validity of the leading order QCD description of the hard scattering
process and to extract information on the subprocesses which
contribute to the observed jet rate.
This will finally allow the determination of parton densities in the photon
as a function of $x_\gamma$.
The main goal is to derive, for the first time, the gluon momentum
distribution $x_\gamma g(x_\gamma)$ in the photon over a large
range in $x_\gamma$.

The correlation between true and reconstructed parton
momenta is shown in Fig.\ref{correlation}d.
The migration effects are large, so       the unfolding procedure -
described in Section 5 - is used to correct from the
reconstructed $x_\gamma^{vis}$ to the true $x_\gamma$ distribution.
As explained before, the unfolding
enforces agreement with the $x_\gamma^{vis}$ distribution (Fig.\ref{2jet}a),
whereas the agreement of the reweighted distributions of other variables
simulated by the Monte Carlo model with the actually observed data
distributions (Fig.\ref{2jet}b-d) delivers important checks of the
transformation:
\begin{figure}[tbp]
\begin{center}
\epsfig{file=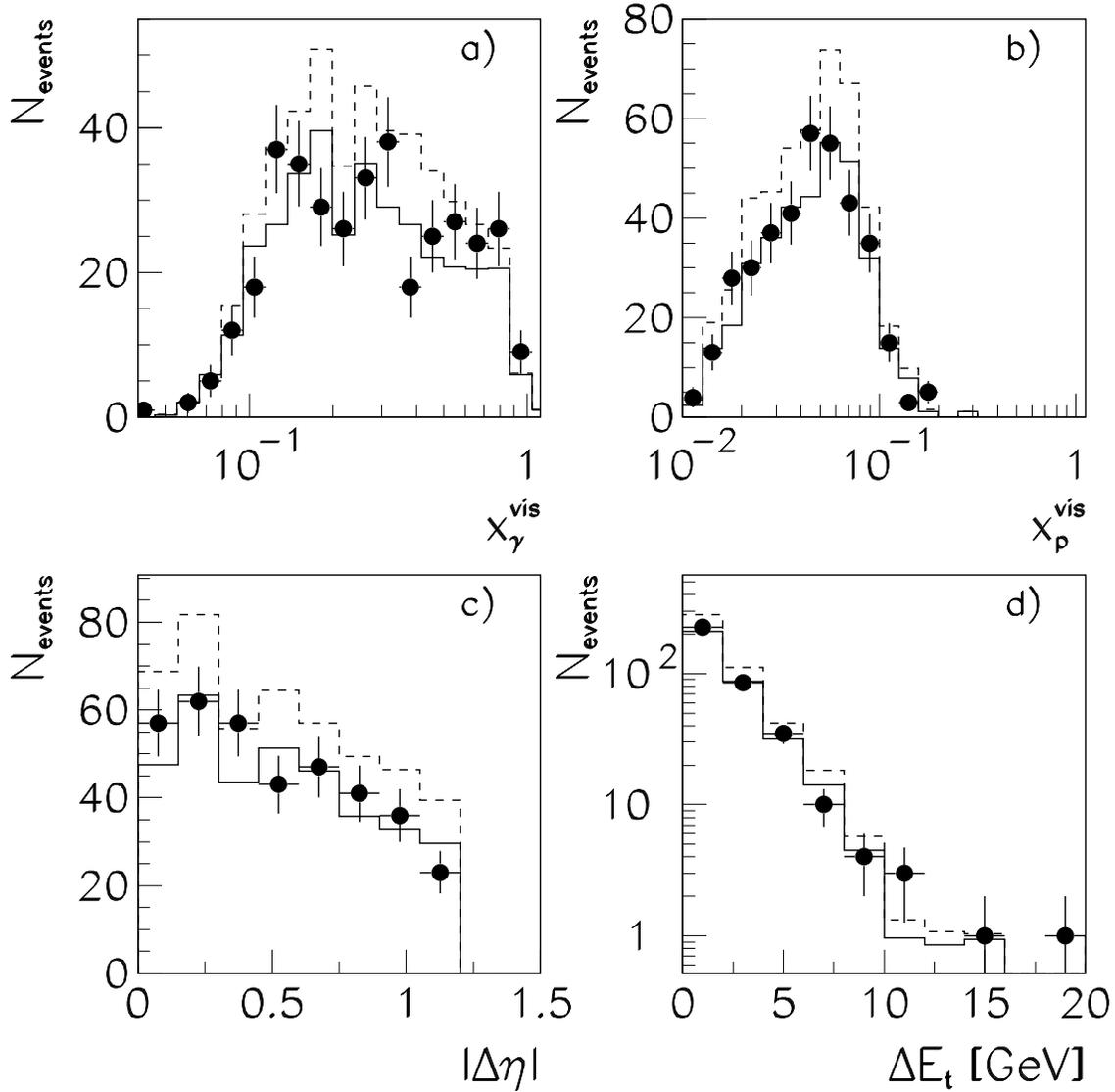,width=15cm}
\end{center}
\caption [$2jet] {\label{2jet} Distributions of the
2-jet data (circles) are compared with the PYTHIA calculations
before (dashed line: absolute prediction) and after (full line)
the reweighting procedure:
a) momentum fraction of the parton from the photon $x_\gamma^{vis}$,
b) momentum fraction of the parton from the proton $x_p^{vis}$,
c) difference in the jet rapidity $|\eta^{jet1}-\eta^{jet2}|$
   which is a measure of the scattering angle in the parton-parton system,
   and
d) the difference in the jet transverse energy $|E_t^{jet1}-E_t^{jet2}|$.}
\end{figure}
\begin{enumerate}
\item Fig.\ref{2jet}b) the fraction of the parton momentum from the
      proton side was determined using
      $x_p^{vis}=
0.5 \left(E_t^{jet1}e^{\eta^{jet1}}+E_t^{jet2}e^{\eta^{jet2}}\right)/E_p$.
      For most of the events $x_p^{vis}$ is above 0.01
      where the quark content of the proton is well established
      by lepton-nucleon scattering experiments and the gluon
      content is known to a level of $15\%$.
%     The agreement between data and calculation shows that the parton
%     distributions in the proton are known with sufficient accuracy.
\item Fig.\ref{2jet}c) the difference in the rapidities of the jets is
      related to the scattering angle $\Theta^*$ of the partons in the
      CM frame of the parton-parton scattering process:
      $\tanh(|\eta^{jet1}-\eta^{jet2}|/2)=\cos \Theta^*$.
      The distributions of $\cos \Theta^*$ are predicted
      by LO QCD for all combinations of the interacting
      partons.
      This is a basic QCD prediction which has to be fullfilled.
\item Fig.\ref{2jet}d) the differences in the transverse energies
      of the jets are large compared to the experimental resolution
      and reflect large average transverse momenta of the incident
      partons.
      Initial state radiation effects are important and to a smaller
      extent large transverse energies of the photon remnant.
\end{enumerate}
The distributions in Fig.\ref{2jet}b-d
demonstrate that a leading order Monte-Carlo
model for the hard scattering process is able to give a consistent
description of the observed 2-jet events by adjusting only the photon
structure function.
Higher order effects as described by initial state parton showers
and multiple interactions to describe the enhanced energy flow at
large rapidities are however essential to describe the data.

\begin{figure}[tbp]
\begin{center}
\epsfig{file=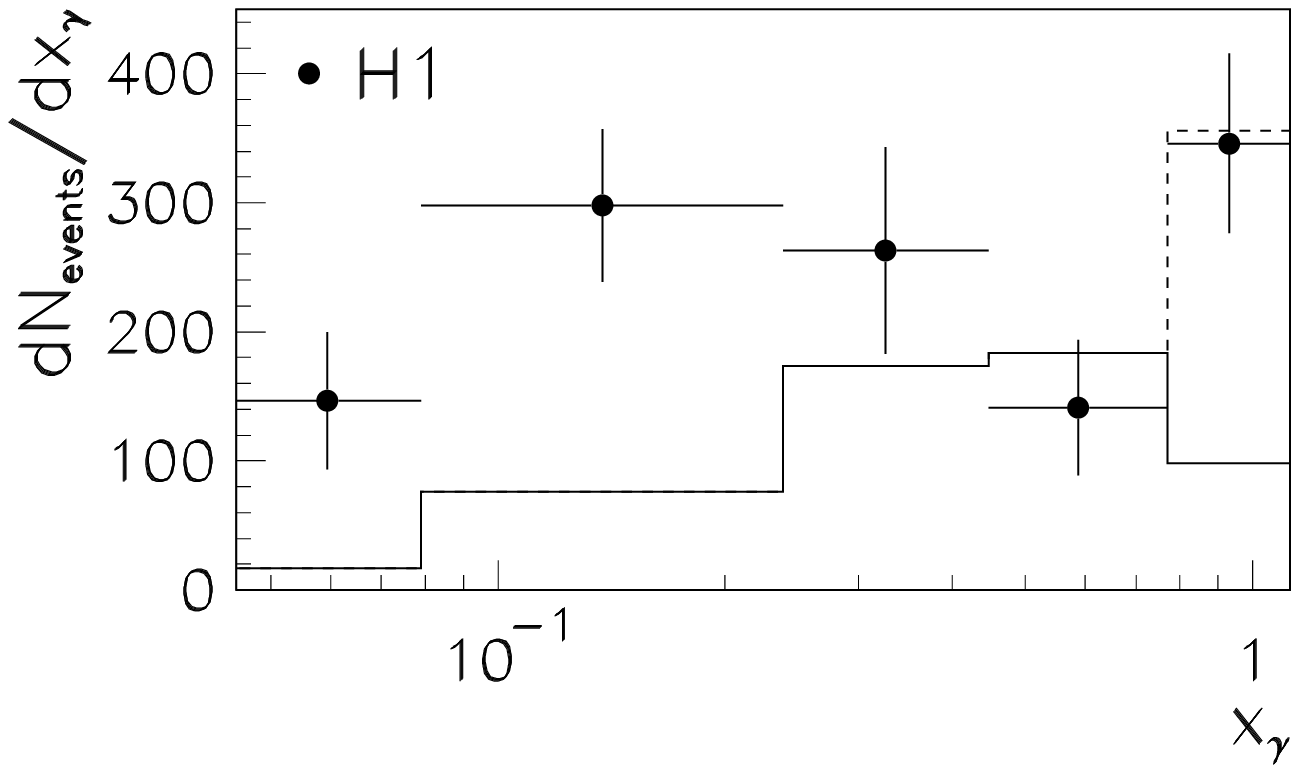,width=16cm,height=10cm}
\end{center}
\caption [$xgamma] {\label{xgamma}
2-jet event distribution (points) of the true fractional
momentum $x_\gamma$ of the parton from the photon.
Only the statistical errors are shown.
The full line represents the predicted contribution
of the quark resolved photon processes, the dashed line shows
the size of the direct photon calculation as obtained by the
PYTHIA Monte Carlo.}
\end  {figure}

Fig.\ref{xgamma} shows the unfolded $x_\gamma$ distribution
compared to calculations of the PYTHIA generator.
The full line represents the prediction of the resolved photon
processes with a quark on the photon side at the hard parton-parton scattering
process.
The quark distribution in the photon as determined by
two photon experiments is input in the form of the GRV-LO
parametrization ~\cite{ggrv}.
The quark induced processes contribute mainly to the central region
of the distribution corresponding to parton momenta around $x_\gamma=0.5$.

The prediction of the direct processes are shown as a dashed line
above $x_\gamma\ge 0.77$.
They account for 3/4 of the events in the highest $x_\gamma$ bin.
Together with the quark component the size of the direct photon calculation,
using the GRV-LO parametrization of the parton densities in the proton,
is consistent with the data at large $x_\gamma$.

At small parton momenta $x_\gamma\le 0.2$ the contribution
of  events initiated by a quark from the photon as predicted by
the given model is  clearly below that seen in the data.
This  suggests explaining the additional events at small
$x_\gamma$ by gluons entering the hard process from the photon side.

\begin{figure}[tbp]
\begin{center}
\epsfig{file=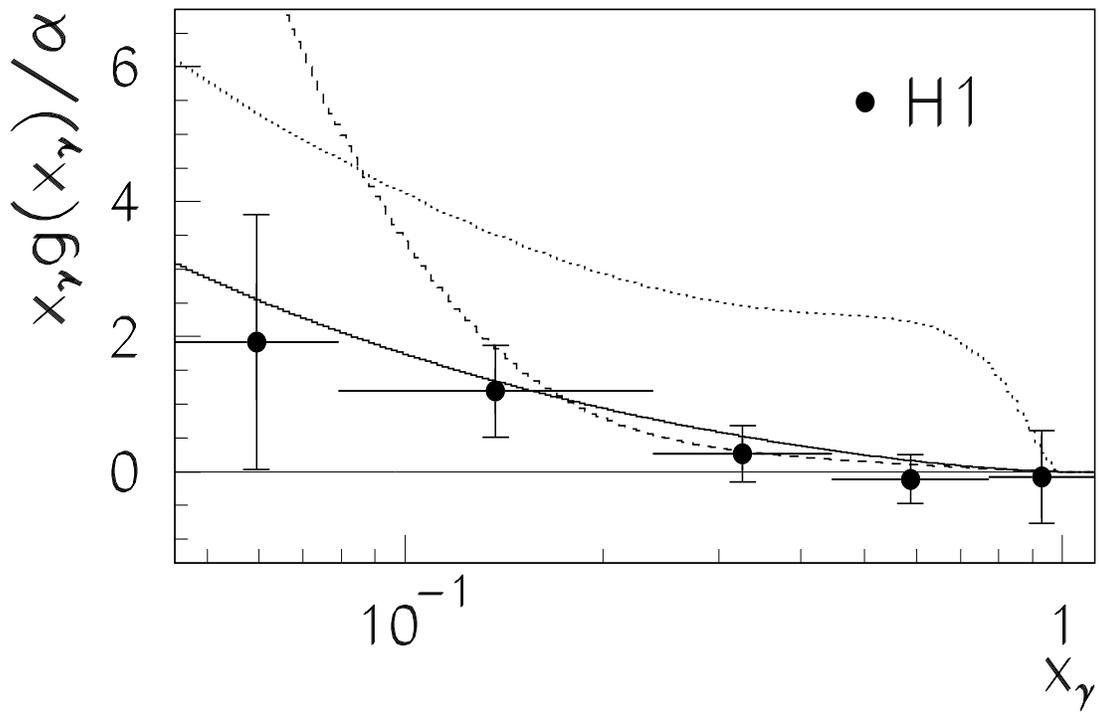,width=16cm,height=10cm}
\end{center}
\caption [$gluon] {\label{gluon}
The gluon density of the photon divided by the
fine structure constant $\alpha=1/137$ (data: circles)
at the scale $<p_t>^2=75$ GeV$^2$.
For comparison the GRV-LO (full) and the LAC1 (dashed)
and the LAC3 (dotted) parametrizations are shown.}
\end  {figure}

%The gluon momentum distribution can be determined from this
%measurement by subtracting the predicted direct and resolved quark
%contributions from the data.
By subtracting the predicted direct and resolved quark contributions from
the data the event distribution of the gluon contribution remains.
Comparison with the distribution calculated for
processes initiated by a gluon from the photon  side,
yields correction factors to be applied to the
modelled distribution.
%The comparison with the events of the calculation, where a gluon enters
%the parton-parton scattering process from the photon side, gives correction
%factors for the modelled gluon distribution.
The resulting gluon density is shown in Fig.\ref{gluon}
(see also Table~\ref{parton_values}):
the total errors include statistical and systematic errors
where all contributions have been added in quadrature.
The systematic errors were calculated from the following sources:
\begin{enumerate}
\item \label{escale}
Error in knowledge of the calorimeter energy scale of $\pm 5\%$.
%\item \label{etrig}
%The error in the determination of the trigger efficiency of $\pm 1\%$,
\item \label{pedcor}
The statistical error in the determination of the hadronic pedestal
correction of $\pm 10\%$.
\item \label{quark}
Uncertainty in the quark density of the photon estimated
      conservatively as $\pm 30\%$.
\end{enumerate}
The systematic errors are dominated at small $x_\gamma$ values
by errors \ref{escale} and \ref{pedcor}, whereas at high values of $x_\gamma$
error \ref{quark} is as large as errors \ref{escale} and \ref{pedcor}.
An additional error from
                        the luminosity measurement of $5\%$ is not included in
the Fig.\ref{gluon}.

To check whether or not a gluon content in the photon is needed to explain
the observed rate in the data, the following test was carried out using
the $x_\gamma^{vis}$ distribution:
the observed data rate in Fig.\ref{xgamma}
was reduced by lowering the energy scale of the
calorimeter according to the error source \ref{escale}.
The remaining sample was further reduced by subtracting more energy
from each jet according to error source \ref{pedcor}.
The Monte Carlo prediction for processes with a quark from the photon however
was raised according to the uncertainty given in item \ref{quark} above.
With the assumption that no gluon exists in the photon
the remaining data events were compared in a $\chi^2$ test
to the predicted number of quark and direct events of the Monte Carlo
calculatio
giving a probability below $0.1\%$ for this hypothesis.
Thus, a gluon contribution in the photon is needed to explain the observed data

The average parton transverse momentum of the selected events is
$<p_t>^2=75$ GeV$^2$ where $p_t$ is used as factorization
and renormalization scale for the QCD calculation.
A change of $\Lambda$ from $200$ MeV to
$300$ MeV reduces the parton density by $30\%$.
%Variations of the factorization scale using the QCD evolution equations
%have a negligable influence on the result.
A variation of the gluon density in the proton by $\pm 15\%$ \cite{nmc}
changes the gluon density in the photon by $\mp 15\%$.
%Ignoring the effect of the jet pedestal correction explained in Section 4,
%the gluon density increases at the two lowest $x_\gamma$ points by
%one standard deviation of the total errors.

The gluon density in the photon in Fig.\ref{gluon}
is compared to the parametrizations of
GRV-LO\cite{ggrv}, LAC1 and LAC3\cite{lac}.
A high gluon density at large parton momenta, as suggested by
the LAC3 parametrization, is clearly excluded.
This is consistent with previous observations at HERA \cite{H1old},
TRISTAN and LEP \cite{gamgam}.
Above $x_\gamma\ge 0.08$ both the GRV-LO and LAC1 parametrizations
of the gluon distribution are consistent with the data.
The strong rise of the LAC1 parametrization below $x_\gamma\le 0.08$
is not supported, while the GRV-LO distribution is consistent
with the data.
In the latter model the gluon distribution of the photon
is essentially generated by QCD radiation.

\section{Conclusion}

Photoproduction processes with at least two high $E_t^{jet}$ jets
in the final state were studied with the H1 detector.
The jet quantities as observed in the detector were unfolded to
leading order parton quantities using
jet-parton correlations based on a specific version of
the PYTHIA QCD generator.
Within the present experimental errors the data are --
after a correction of the jet energy -- consistent
with this leading order description of the hard scattering process
together with higher order processes in the initial and final state
and multiple interactions, i.e.
interactions between partons of the photon and proton spectators.

Single inclusive differential $ep$ parton cross sections were derived
as a function of the transverse momentum $p_t\ge 7$ GeV and pseudo-rapidity
$0\le\eta\le 2.5$ of the partons
for photon energies corresponding to $0.25\le y\le 0.7$ and momentum
transfer below $Q^2\le 0.01$ GeV$^2$.
They agree well with QCD predictions.
A direct study of 2-jet kinematics was used to determine
the momentum fraction of the partons from the photon.
At large $x_\gamma$ it is
consistent with the predictions of the direct component plus
%that part of
the resolved component initiated by quarks from the photon.
At low parton momentum fraction the observed jet rate in the data can only be
explained by a gluon component in the photon.
For the first time
a leading order gluon distribution was derived down to $x_\gamma=0.04$.
%Here, for example,
%the gluon content in the photon is several hundred times smaller
%than that in the proton.
The average scale was $<p_t>^2=75$ GeV$^2$ corresponding to
the mean transverse momentum squared of the final state partons.
Gluon distributions with high density at large $x_\gamma$ or
steeply rising gluon distributions at small $x_\gamma$ are disfavoured.

\begin{table}
%p-transversal
\begin{center}
\begin{tabular}[h]{|c||c|c|c|c|}
\hline
      $p_t$ [GeV]             & 8.2  & 10.5  & 13.9  & 19.6  \\
     \hline\hline
      $d\sigma/dp_t$ [nb/GeV] & 3.57 & 0.90  & 0.31  & 0.049 \\
     \hline
      $\sigma_{stat}$         & 0.19 & 0.09  & 0.05  & 0.013 \\
     \hline
      $\sigma_{syst}$         & 1.52 & 0.32  & 0.13  & 0.018 \\
     \hline
\end{tabular}
\end{center}
% Rapidity
     \begin{center}
     \begin{tabular}[h]{|c||c|c|c|c|}
     \hline
      $\eta$                  &  0.42 & 1.25 & 2.08  \\
     \hline\hline
      $d\sigma/d\eta$ [nb]    &  5.61 & 5.25 & 3.50  \\
     \hline
      $\sigma_{stat}$         &  0.37 & 1.02 & 0.16  \\
     \hline
      $\sigma_{syst}$         &  2.10 & 2.07 & 1.60  \\
     \hline
     \end{tabular}
     \end{center}
% gluon
\begin{center}
\begin{tabular}[h]{|c||c|c|c|c|c|}
\hline
      $x_\gamma$                    & 0.059 & 0.14  & 0.33 &  0.59 &  0.93 \\
     \hline\hline
      $x_\gamma g(x_\gamma)/\alpha$ & 1.92  & 1.19  & 0.26 & -0.12 & -0.08 \\
     \hline
      $\sigma_{stat}$               & 0.87  & 0.34  & 0.24 &  0.15 &  0.61 \\
     \hline
      $\sigma_{syst}$               & 1.68  & 0.59  & 0.33 &  0.33 &  0.30 \\
     \hline
\end{tabular}
\end{center}
\caption [$parton_values] {\label{parton_values}
Single inclusive differential parton cross sections (Fig.\ref{pT},\ref{eta})
and
the gluon distribution in the photon (Fig.\ref{gluon})
together with their statistical and systematic errors.}
\end{table}

%\newpage
\vspace{1cm}
\par\noindent
{\bf Acknowledgement}

\noindent
We are grateful to the HERA machine group whose
outstanding efforts made this experiment possible. We appreciate the immense
effort of the engineers and technicians who constructed and maintain
the H1 detector. We thank the funding agencies for financial support.
We acknowledge the support of the DESY technical staff. We wish to
thank the DESY directorate for the support and hospitality extended to
the non-DESY members of the collaboration.

\renewcommand{\baselinestretch}{1.0}
{\Large\normalsize}

\end{document}